# Physical insight into reduced surface roughness scattering in strained silicon inversion layers

Chris Bowen and Ryan Hatcher
*Lockheed Martin Advanced Technology Laboratories, Cherry Hill, NJ 08002*

A seemingly anomalous enhancement of electron mobility in strained silicon inversion layers at high sheet densities has exposed a conspicuous gap between device physics theory and experiment in recent years. We show that the root of this discrepancy is due to a bulging effect in the electron $\Delta_4$ wavefunction at the silicon surface. This renders $\Delta_4$ electrons more susceptible to perturbations in surface structure thereby increasing surface roughness scattering for these states. Strain engineering utilized by the CMOS industry reduces the relative occupancy of the $\Delta_4$ states resulting in less overall surface roughness scattering in the channel. We show that the origin of this effect can be explained by moving beyond the effective mass approximation and contrasting the properties of the $\Delta_2$ and $\Delta_4$ wavefunctions in a representation that comprehends full crystal and Bloch state symmetry.

Strain has been used in production CMOS to enhance the performance of both the NMOS and PMOS transistors for the past decade [1]. In the case of the NMOS transistor, a clear understanding of this strain induced enhancement has remained elusive [2]. A detailed and thorough calculation using effective mass Hamiltonians was conducted by Fischetti who pointed out just how dire the situation was [2]. An important conclusion of his work was that strained silicon inversion layers should be expected to exhibit increased surface roughness scattering [2]. And since surface roughness scattering limits mobility in strong inversion, the strain employed by the CMOS industry should be degrading the NMOS transistor when in reality the opposite is true [1-12]. While an impressive density functional theory (DFT) based simulation has shown the ability to reproduce experimental data in the strong inversion regime, little physical insight has been gleaned from these calculations [13]. The purpose of this work is to provide insight into the fundamental physics which leads to reduced surface roughness scattering in strained silicon inversion layers.

In (100)-oriented silicon inversion layers, the six-fold degeneracy of the conduction band at $\Delta$ is lifted into four in-plane ($\Delta_4$) and two normal ($\Delta_2$) valleys [19]. This results in an increased distribution of carriers into subbands derived from the $\Delta_2$ valleys which possess a smaller in-plane effective mass [12,18]. The application of in-plane tensile or normal compressive stress breaks the $\Delta$-degeneracy in the same direction as the inversion layer field resulting in a further increased occupation of the $\Delta_2$ subbands [12,18]. The question we seek to answer is how this increase in $\Delta_2$ occupancy results in reduced surface roughness scattering.

Electron $\Delta_2$ and $\Delta_4$ wavefunctions are shown in Figure 1 for an unstrained silicon inversion layer calculated using the Socorro code [14] with norm-conserving pseudopotentials [15], a plane wave basis set, and the generalized-gradient approximation (GGA) for exchange and correlation [16]. The plane wave cutoff energy is 408 eV and the Brillouin zone was sampled using the Monkhorst-Pack technique [17]. The silicon surface is hydrogen passivated and a constant electric field of $10^6$ V/cm is applied normal to the surface. In striking contrast to effective mass calculations [2], DFT predicts a dramatic increase in the $\Delta_4$ probability density at the silicon surface despite the delocalized nature of the higher energy $\Delta_4$ state. The $\Delta_4$ wavefunction punches through both of the $\Delta_2$ wavefunctions resulting in an increased sampling of the surface for electrons occupying $\Delta_4$ states. This will clearly result in a dramatic increase in scattering due to any surface structural perturbation for these electrons. Therefore, strain-induced reduction of the $\Delta_4$ occupancy will result in a reduction of channel surface roughness scattering. We propose this to be the underlying physical phenomenon which resolves the discrepancy between experiment and calculations performed with the effective mass approximation for the electron wavefunction. We note here that we have also observed this effect at Si/SiO$_2$ interfaces. Since the goal of this letter is to convey the fundamental physics behind this phenomenon, we choose to study the simpler hydrogen passivated surface.

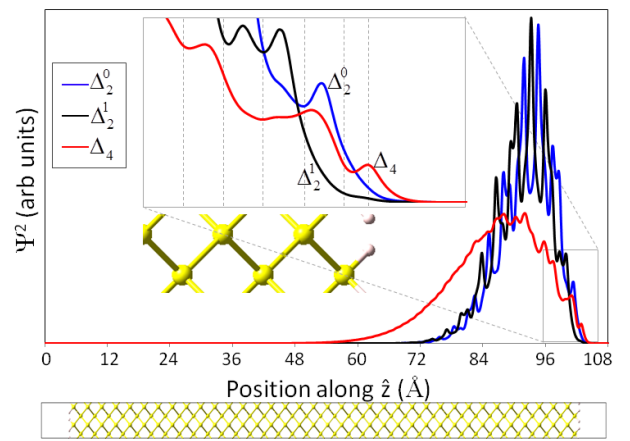

**FIG 1.** The planar average of $|\psi|^2$ for the $\Delta_2$ and $\Delta_4$ states along the direction of quantization for a thin Si(001) slab under a bias of $10^6$ Vcm$^{-1}$. The slab is 96Å thick and H-terminated at both surfaces. The inset reveals details of the wavefunctions near the surface. Note the $\Delta_4$ density punching through the $\Delta_2$ densities at the surface

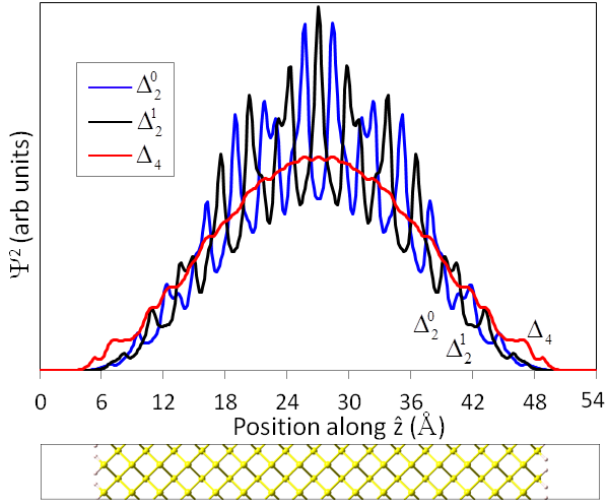

**FIG 2.** The planar average of $|\psi|^2$ for the $\Delta_2$ and $\Delta_4$ states along the direction of quantization a thin Si(001) slab with no external bias. The slab is 42Å thick and H-terminated at both surfaces. The bulging effect in the $\Delta_4$ density at the surface is more pronounced here compared to FIG 1 because the well is symmetric resulting in a narrower spectrum in reciprocal space for the ground state.

To more clearly describe what causes this density bulging at the surface, we consider the $\Delta_2$ and $\Delta_4$ ground states in a 4nm hydrogen passivated silicon quantum well with no applied field. As seen in Figure 2, the magnitude of the $\Delta_4$ wavefunction is larger than that of the $\Delta_2$ state at the surface in qualitative agreement with Figure 1. Decomposing the $\Delta_4$ ground state into its orbital components, we find that this wavefunction contains both even and odd envelopes (Figure 3a) [20, 21]. Both envelopes are superpositions of Bloch states with $k=\pi/L$ (where $L$ is the well width) and therefore the odd envelope is maximum *at the surface*. The resulting probability distribution of the $\Delta_4$ ground state contains both $\cos^2((z-z_0)\pi/L)$ and $\sin^2((z-z_0)\pi/L)$ components (where $z_0$ is the center of the well) the latter of which is indicated with an overlay of $\sin^2((z-z_0)\pi/L)$ on a pair of the odd envelopes in Figure 3a. Returning now to Figure 2, this characteristic is evident in the fully spatially resolved plane wave DFT wavefunction, and it is these odd envelopes that are the root cause of the surface bulging effect of the $\Delta_4$ state.

Though previously discussed by Boykin [20,21], the existence of both even and odd parity envelopes in a ground state may seem somewhat counterintuitive and merits some discussion. We first note that this effect isn't comprehended by the effective mass approximation and this is ultimately the reason behind the qualitative discrepancy between such calculations and experiment [2]. The explanation, however, is straightforward from the perspective of an orbital representation of the wavefunction. The odd envelopes of the ground state arise from in-plane orbital Bloch sums that possess odd parity in the dimension of quantization. As illustrated in Figure 4 odd parity orbitals must be modulated by an odd parity envelope to preserve the overall even parity of the ground state wavefunction in a symmetric confining potential. Returning to Figure 3a, the Bloch sums consisting of $p_y$, $p_z$, $d_{xy}$, $d_{xz}$ orbitals are all odd in the z-direction and thus exhibit odd envelopes.

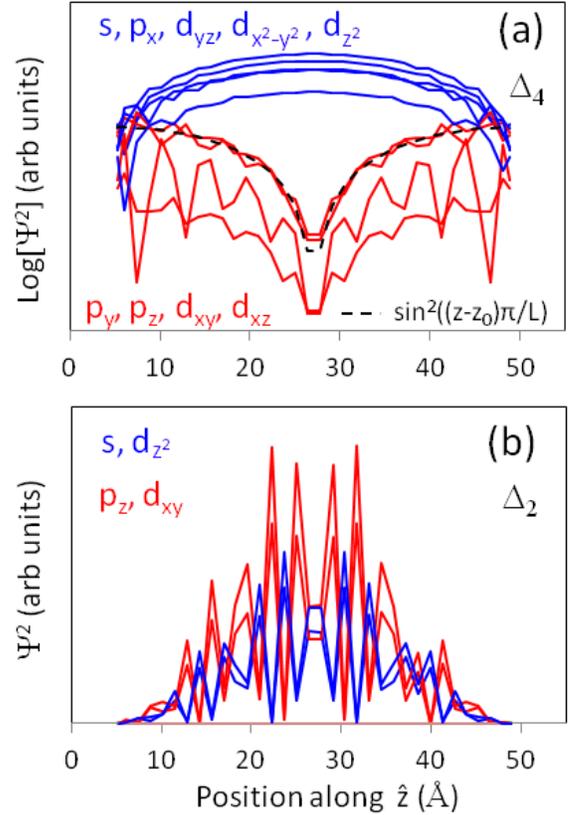

**FIG 3.** (a) Orbital decomposed envelopes of the $\Delta_4$ ground state wavefunction in a 4nm hydrogen passivated silicon quantum well computed with DFT. In-plane orbital Bloch sums that are odd in the quantization dimension result in odd parity envelopes causing a bulging of the wavefunction at the silicon surface. The wavefunctions are plotted on a log scale to reveal the functional form of the odd envelopes. (b) Same plot for the $\Delta_2$ ground state which also exhibits opposite parity for even and odd orbital components. The rapid oscillatory behavior of these envelopes suppresses any bulging effect.

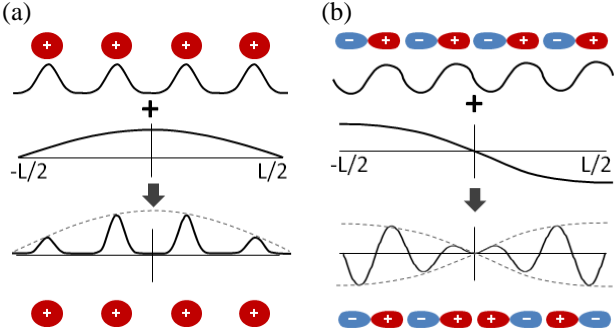

**FIG 4.** *Ground state of a four atom quantum well. For even basis functions such as shown in a), the envelope must be even in order to maintain even parity of the state. For odd basis functions such as one shown in b), the envelope must be odd in order to maintain an even parity. Since the envelopes correspond to $k=\pi/L$, any contribution from even orbitals is suppressed by the modulation at the surface while any contribution from odd orbitals is maximal at the surface.*

To express this effect in a more general manner, consider an arbitrary state in the basis of the Bloch states of an infinite 1D periodic potential

$$\Psi = \int_{BZ} dk \sum_n c_{nk} \psi_{nk} \quad (1)$$

Where each Bloch state is constructed from a planewave and a Bloch function, $u_{nk}$, which shares translational symmetry with the periodic potential (e.g. $u_{nk}(z+ma_0)=u_{nk}(z)$ for any integer $m$)

$$\psi_{nk} = u_{nk} e^{ikz} \quad (2)$$

The Bloch state is unique for any $k = -\pi/a_0 \ldots \pi/a_0$, where $a_0$ is the length of the unit cell of the periodic potential. The Bloch states can be decomposed by parity about $z=0$.

$$\psi_{nk} = \psi_{nk}^{even} + \psi_{nk}^{odd} \quad (3)$$

Substituting (3) into (2) and decomposing the Bloch functions, $u_{nk}$ by parity yields

$$\psi_{nk}^{even} = u_{nk}^{even} \cos kz + i u_{nk}^{odd} \sin kz$$
$$\psi_{nk}^{odd} = u_{nk}^{odd} \cos kz + i u_{nk}^{even} \sin kz \quad (4)$$

as can be seen above, for any parity decomposed Bloch state (i.e. $\psi_{nk}^{even}$ or $\psi_{nk}^{odd}$), there is a constant $\pi/2$ phase shift between the contributions from the even and odd components of $u_{nk}$. This phase shift corresponds to a real space phase separation of $\pi/2k$, and can have an appreciable effect on the density distribution for small values of momentum.

For example, consider an infinite 1D square well potential of length $L$ consisting of 8 periodic unit cells with period $a_0$. Suppose there exists an eigenstate, $\Psi$, equal to the sum of two Bloch states

$$\begin{aligned}\Psi(z) &= \tfrac{1}{\sqrt{2}}\left(u_k e^{ikz} + u_{-k} e^{-ikz}\right) \\ &= \sqrt{2}\left(u_k^{even} \cos kz + i u_k^{odd} \sin kz\right)\end{aligned} \quad (5)$$

Where $u_{\pm k}$ are Bloch functions of the potential unit cell and it has been assumed that the solution is even so that $u_k^{even} = u_{-k}^{even}$ and $u_k^{odd} = -u_{-k}^{odd}$. The density is given by

$$n(z) = 2\left|u_k^{even}\right|^2 \cos^2(kz) + 2\left|u_k^{odd}\right|^2 \sin^2(kz) \quad (6)$$

If $k$ is small, then the spatial separation of the phases ($\pi/2k$) is large resulting in significant density bulging at the edge of the well as shown in Figures 5b and 5c. The momentum dependence of the spatial separation suggests that the surface bulging effect will be more pronounced in bound states comprised of a narrow spectrum of Bloch wavevectors in the quantization direction. And indeed comparing Figures 1 & 2 it is evident that the $\Delta_4$ surface bulging is more pronounced for the symmetric well than the field confined triangular well. A Fourier analysis of these wavefunctions reveals a 25% broadening of the k-space spectrum for the field confined $\Delta_4$ state commensurate with the expected trend.

We now must address the question as to why this surface bulging does not occur in the $\Delta_2$ wavefunctions despite the fact that these states also have contributions from odd orbital Bloch sums just like the $\Delta_4$ states. The difference is due to the $\pi/2k$ dependence of the real space separation of the phases. Specifically, in the quantization direction, the $\Delta_2$ wavefunctions are superpositions of Bloch states with phases separated by $\sim a_0/2$ whereas the $\Delta_4$ state is comprised of Bloch states with phases separated by $\sim L/2$ (where $L=ma_0$ with $m$ periodic layers of crystal in the well) [22,23]. This is clearly illustrated in Figure 3b, which contains the even and odd orbital envelopes of the $\Delta_2$ ground state and Figure 5d which compares the density distribution for one solution with a small wavevector (similar to the $\Delta_4$ state) and another with a large wavevector (similar to the $\Delta_2$

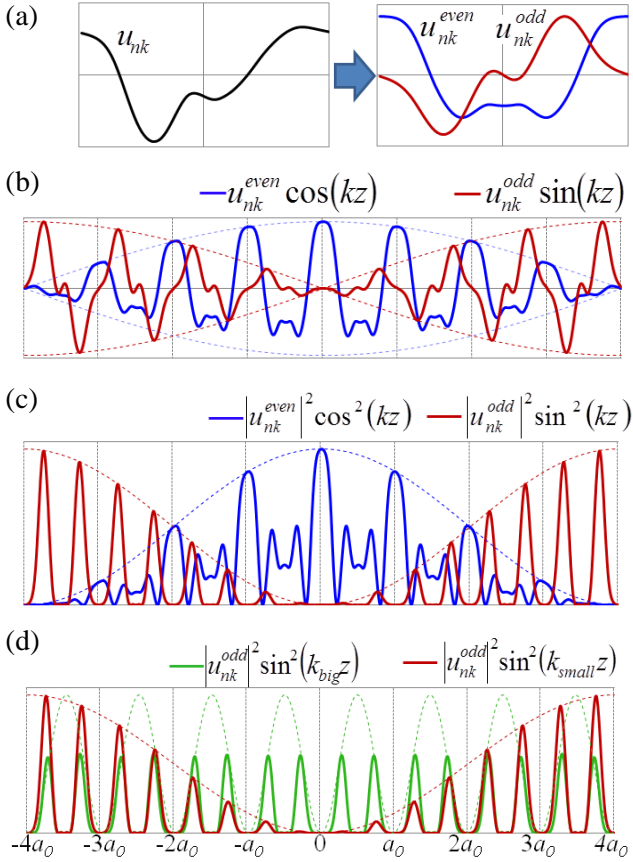

**FIG 5.** *For a periodic potential with unit cells of length $a_0$, an even wavefunction is constructed from two Bloch states of the infinite periodic potential with $\pm k$ and a Bloch function $u_{nk}$. For such a wavefunction there is a $\pi/2$ phase shift between the (a) even and odd components of the Bloch function. The spatial extent of this phase shift is inversely proportional to $k$ as can be seen in the contributions to both the wavefunction (b) and density (c) where $k=\pm\pi/8a_0$ with the corresponding real space phase shift of $4a_0$. (d) For a wavefunction constructed from Bloch states with large momentum, $k_{big} = \pm\pi/a_0$, the real space phase separation is small as is the redistribution of density – especially compared to a wavefunction constructed from Bloch states with small momentum, $k_{small} = \pm\pi/8a_0$*

states).

Comparing the $\Delta_2$ wavefunctions in Figure 1, it is readily apparent that the higher energy state is more suppressed from the surface. We find that this occurs whenever inversion symmetry is broken resulting in a non-negligible energy splitting (single digit meV) of the $\Delta_2$ states [22,23]. This effect is noteworthy because roughly half of the electrons residing in $\Delta_2$ are exceedingly isolated from the silicon surface.

While the plane wave DFT calculations presented above demonstrate very nicely the physics behind the $\Delta_4$ surface bulging effect, these calculations are numerically expensive and a reduced representation is desirable. In order to capture the essential physics, a wavefunction representation that includes the full crystal and Bloch state symmetry properties is required. Fortunately, the empirical tight binding method (ETBM) comprehends these properties using a basis set that allows for the simulation of structures comprised of up to tens of millions of atoms [24,25]. In fact, we first noticed the $\Delta_4$ surface bulging effect in calculations using this basis. In Figure 6 we show a comparison between the orbital-decomposed plane wave DFT and ETBM wavefunctions (generated with the NEMO5 simulation tool [26]) for the 4nm silicon quantum well. While the ETBM calculation qualitatively captures the physics it underestimates the magnitude of the effect. The reason for this is that the relative orbital contributions of the ETBM wavefunction do not match those of the DFT calculation. This is not surprising because the weighting of wavefunction orbital contributions was not taken into consideration when the Hamiltonian was parameterized. This is a situation which begs for ETBM parameterizations that are informed by ab-initio wavefunctions [27,28]. The exercise of parameterizing tight binding Hamiltonians that result in quantitative agreement with DFT in order to conduct detailed mobility calculations that include the physics presented in this letter is left to a future publication.

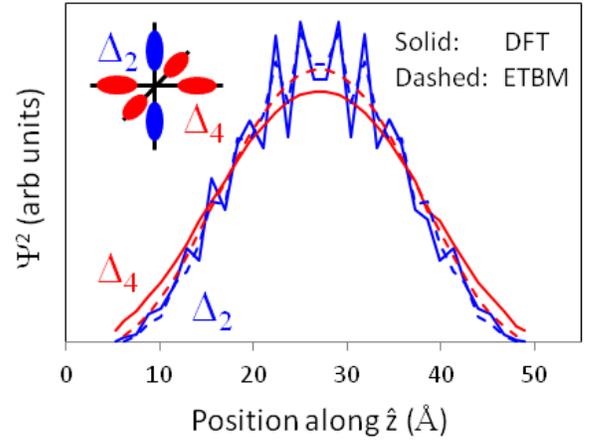

**FIG 6.** *Trace over the orbital coefficients for spherical harmonic decomposed DFT (solid lines) and ETBM (dashed lines) wavefunctions for the $\Delta_2$ (blue lines) and $\Delta_4$ (red lines) ground state wavefunctions.*

We have shown that electron $\Delta_4$ states in silicon inversion layers are subject to increased surface roughness scattering due to a bulging effect of the wavefunction at the silicon surface. Although the same underlying physics is present in the $\Delta_2$ states, the bulging effect is effectively suppressed due to the large crystal momentum (short real space wavelength) in the quantization direction which in turn minimizes any spatial redistribution of density. In-plane tensile or normal compressive strain acts to depopulate $\Delta_4$ states and will therefore result in reduced surface roughness scattering in silicon NMOS transistors. The bulging effect is a direct manifestation of the underlying crystal and Bloch state symmetry properties of silicon. It follows that models must comprehend these properties in order to be predictive. While the empirical tight binding method includes the requisite physics, parameterizations that reproduce the correct wavefunction orbital weighting must be used to obtain quantitatively accurate calculations. We also believe the insight that $\Delta_2$ and $\Delta_4$ electrons interact with the surface very differently can be leveraged to develop improved mobility models for Boltzmann transport and drift-diffusion based device simulation.

The authors acknowledge helpful conversations with Timothy Boykin and Borna Obradovic.